\documentstyle[12pt,a4wide]{article}

\input epsf

\def\slp{p\hspace{-0.45em}{ /}}

\title{
\begin{flushright}
{\normalsize DO--TH 94/16}
\end{flushright}
\vspace{2cm}
$\cal CP$ violation in the time evolution \\
of the decay $K^{0} \rightarrow \pi^{0} e^{+} e^{-}$
\thanks{ This work was supported in
part by the Bundesministerium fr Forschung und Technologie, 05-6D093P,
Bonn, FRG and by the CEC Science Project $\mbox{n}^{o}$ SC1-CT91-0729.}
}
\author{ Gerhard O. Khler, Emmanuel A. Paschos \\
Institut fr Physik, Universitt Dortmund \\
D -- 44221 Dortmund, Germany }
\begin{document}
\maketitle
\begin{abstract}
We examine the possibility of extracting $\cal CP$-violating terms in the
decay \\ $K^{0} \rightarrow \pi^{0} e^{+} e^{-}$ by studying the time
evolution of a $K^{0}$ beam.
We focus on the interference region and search for clear effects.
We find that experiments which average over the electron and positron
momenta can detect $\cal CP$ violation as an oscillation in the decay
rate. The branching ratio is $( 3 - 5 ) \times 10^{-12}$ and direct
$\cal CP$ violation dominates over a wide range of the parameters.
\end{abstract}
\newpage
\section{Introduction}
One property of the standard model which is still under active
consideration is the origin of $\cal CP$ violation.
Up to now $\cal CP$-odd contributions have been observed only in the
decays of $K^{0}$ mesons \cite{Kaon}.
In the decay one distinguishes two types of $\cal CP$ violation effects:
direct $\cal CP$ violation occuring in the amplitudes ( described by
$\epsilon^{'}$ ) and $\cal CP$-asymmetric terms in the mass matrix which
is called indirect ( described by $\epsilon$ ).
The value of $\epsilon$ is precisely known
(~$|\epsilon| = 2.258 \times 10^{-3}$~), but there are still
uncertainties concerning the value of $\epsilon^{'}$.
The CERN experiment NA31 found \cite{Burkhardt}
\begin{equation}
Re\Big(\frac{\epsilon^{'}}{\epsilon}\Big)\;=\;
( 2.3 \pm 0.7 ) \times 10^{-3}
\end{equation}
while the measurement of the FERMILAB experiment E731 is \cite{Gibson}
\begin{equation}
Re\Big(\frac{\epsilon^{'}}{\epsilon}\Big)\;=\;
( 0.74 \pm 0.59 ) \times 10^{-3}
\end{equation}
which is still consistent with the predictions of the superweak theory.
Thus it is still interesting to investigate other processes in order
to find if direct $\cal CP$ violation is different from zero and
providing another crucial test of the standard model.
Examples for such processes are B meson and rare $K^{0}$ meson decays,
which are actively investigated both theoretically and experimentally.
A promising decay channel is $K^{0} \rightarrow \pi^{0} e^{+} e^{-}$,
where $\cal CP$ violation may be relatively large.
The specific decay $K_{L} \rightarrow \pi^{0} e^{+} e^{-}$ has been
studied extensively and its status was recently reviewed
\cite{Winstein-Wolfenstein,Rosner}.

We study the time development of this decay channel
starting with a pure $K^{0}$ beam and pose the
question if one could identify a $\cal CP$-violating signal in the
interference region. In particular, we are interested in a signature of
direct $\cal CP$ violation which we present in this article.
We will show that an experiment which studies the time development of
$K^{0}$ decays and averages over the momenta of electron and positron is
sensitive to $\cal CP$-violating terms. The new effect appears in the
interference region of the $K_{S}$ and $K_{L}$ component and manifests
itself as a time oscillation. The effect follows from general symmetry
considerations as is explained in the next section. In addition, we
present an estimate for the magnitude of the effect. This experiment is
especially suited for laboratories with intense K beams like Brookhaven
\cite{Littenberg}.

Our paper is organized as follows. In section 2 we derive the formula
describing the time evolution of a pure $K^{0}$ state and classify the
different contributions. In section 3 we discuss the calculations
available for the amplitudes and their dependence on the parameters.
Furthermore we give the range of parameters, which is used later on
in the numerical analysis. In section 4 we present the numerical
results for the time evolution of the $K^{0}$ state with special emphasis
on $\cal CP$ violation in the interference region.
Finally, the reder who is interested on the experimental possibility can
study section 2 and the conclusions in section 4.

\section{Classification of the various amplitudes}
The time evolution of a pure $K^{0}$ state is given in terms of the
time development of the physical states $K_{L}$ and $K_{S}$ as follows
\begin{equation}
|K^{0}(t)> = \frac{1}{\sqrt{2}}\;
	     \Big[e^{- i X_{L} t} (\;|K_{2}> + \epsilon |K_{1}> ) +
	     e^{- i X_{S} t} (\;|K_{1}> + \epsilon |K_{2}> )\Big] .
\end{equation}
The decay proceeds through two intermediate states,
$K^{0} \rightarrow \pi^{0} \gamma$ and
$K^{0} \rightarrow \pi^{0} \gamma \gamma$, with the single or two photons
converting into electron positron pairs.
The decaying kaon has spin 0, and angular momentum conservation demands
the intermediate state $\pi^{0} \gamma$ to be in a p-wave. It follows,
then, that the $\cal CP$ eigenvalue of $\pi^{0} \gamma$ is
$(-1)_{\pi} (+1)_{\gamma} (-1)^{l=1}\;=\;+1$.
Thus $K_{1}$ can decay through the $\pi^{0} \gamma$ channel in terms of
$\cal CP$-conserving parts, whereas $K_{2}$ decays
through the $\cal CP$-violating parts
of the Hamiltonian. We denote these amplitudes as
\begin{eqnarray}
A_{1}&=&< \pi^{0} e^{+} e^{-} |\hspace{0.3cm}{\cal H}_{\gamma}
\hspace{0.3cm}| K_{1} > ; \hspace{1.5cm}{\cal CP}
\mbox{-conserving ( It gives } \\
& &\hspace{6.4cm}{\mbox{indirect }\cal CP}\mbox{ violation through }
\epsilon ) , \nonumber \\
B&=&< \pi^{0} e^{+} e^{-} |\hspace{0.3cm}{\cal H}_{\gamma}
\hspace{0.3cm}| K_{2} > ; \hspace{1.5cm}{\cal CP}
\mbox{-violating ( direct ) .}
\end{eqnarray}
These would be the only two amplitudes if there were no higher order
terms. In fact the decay of a $K^{0}$ can also proceed through the
intermediate state $\pi^{0} \gamma \gamma$, which is higher order in the
electromagnetic coupling. This
contribution to the decay width is not a priori negligible, because, as
we will show, it has to be compared with $\cal CP$-violating terms which
are suppressed. To be specific, the intermediate state of a pion and two
photons has many partial waves so that both ${\cal CP }= + 1$ and
${ \cal CP} = - 1$ states are allowed. Thus the decay
\begin{equation}
K_{2} \rightarrow \pi^{0} \gamma \gamma \rightarrow \pi^{0} e^{+} e^{-}
\end{equation}
is $\cal CP$-conserving with the final state odd under the
$\cal CP$ transformation. In fact the decay $K_{2} \rightarrow \pi^{0}
\gamma \gamma$ has already been observed. We define the relevant
amplitude as
\begin{equation}
A_{2}\;=\;< \pi^{0} e^{+} e^{-} |
\hspace{0.3cm}{\cal H}_{\gamma\gamma}\hspace{0.15cm}| K_{2} > ;
\hspace{1.5cm}{\cal CP}\mbox{-conserving} .
\end{equation}
The decay of a pure $K^{0}$ beam has the general form
\begin{eqnarray}
< \pi^{0} e^{+} e^{-} |\hspace{0.1cm}{\cal H}
\hspace{0.1cm}| K^{0} >( t )
&=& \frac{1}{\sqrt{2}}\;\Big\{e^{- i X_{L} t}\;
\overline{u}(k_{-}) \slp_{K} \Big[ ( B^{+} + \epsilon
A_{1}^{+} + A_{2}^{+} ) + B^{-}
\gamma_{5} \Big] v(k_{+}) \\
& &\hspace{1cm}+\;e^{- i X_{S} t}\;\overline{u}(k_{-}) \slp_{K}
\Big[ A_{1}^{+} + \epsilon ( B^{+} + A_{2}^{+} ) + \epsilon B^{-}
\gamma_{5} \Big] v(k_{+}) \Big\} \nonumber
\end{eqnarray}
where the $+,-$ indicate that the spinors are written out explicitly,
with $A_{1}^{+}, A_{2}^{+}, B^{+}$ being vector amplitudes and $B^{-}$
being the axial-vector. For explicit definitions see equations (12), (17)
and (24).

The term $\epsilon ( B + A_{2} )$ is small in comparison to $A_{1}$ on
several resons:
\begin{itemize}
\item[i)] $B$ is small, being $\cal CP$-violating,
\item[ii)] $A_{2}$ is small, being higher order in electromagnetism
\item[iii)] these two small terms are multiplied by the small parameter
$\epsilon$.
\end{itemize}
Neglecting $\epsilon ( B + A_{2} )$,
the $K_{S}$ decays are $\cal CP$-conserving.
The $K_{L}$ decays contain $A_{2}$, which is $\cal CP$-conserving, and
the $\cal CP$-violating amplitudes $B$ and $\epsilon A_{1}$. The
amplitude $B$ represents direct $\cal CP$ violation, whereas the
violation in $\epsilon A_{1}$ arises through the mass matrix. Since both
terms are very likely suppressed, it becomes necessary to consider the
$A_{2}$ term, as mentioned above.

Next, we compute the time evolution of the decays.
\begin{eqnarray}
\frac{d\Gamma}{ds d\Delta}\;(t)&=& \frac{1}{512\;\pi^{3} m_{K}^{3}}\;
\Big\{\;e^{- \Gamma_{L} t}\;
\Big[ | B^{+} +  A_{2}^{+} + \epsilon A_{1}^{+} |^{2} + | B^{-} |^{2}
\Big]\;+\;e^{- \Gamma_{S} t}\;| A_{1}^{+} |^{2} \nonumber \\
& &\hspace{2.3cm}+\;e^{-\frac{\Gamma_{L} + \Gamma_{S}}{2}\;t}\;2\;
Re\Big[e^{-i \Delta m_{K} \cdot t}\;( B^{+} + A_{2}^{+} + \epsilon
A_{1}^{+} ) A_{1}^{+ *} \Big]\;\Big\} \label{gleichung2.1} \\
& &\hspace{0.5cm}\times\;\frac{1}{2}\;\big[\;
\lambda(s,m_{K}^{2},m_{\pi}^{2}) - \Delta^{2}\;\big] \nonumber
\end{eqnarray}
where
\begin{eqnarray}
s & = & ( p_{K} - p_{\pi} )^{2} \nonumber \\
\lambda( s, m_{K}^{2}, m_{\pi}^{2} ) & = & s^2 + m_{K}^{4} + m_{\pi}^{4}
- 2\;s\;m_{K}^{2} - 2\;s\;m_{\pi}^{2} - 2\;m_{K}^{2} m_{\pi}^{2}
\hspace{1cm}\mbox{and} \nonumber \\
\Delta & = & ( p_{K} - k_{-} )^{2} - ( p_{K} - k_{+} )^{2} . \nonumber
\end{eqnarray}
This expression shows three time intervals: decays for $K_{S}$-,
$K_{L}$ mesons and an interference region. The first two show the typical
exponential behavior for the decays, and the interference has
an oscillatory term as well. We
point out an important property of the interference term. The $A_{2}$
amplitude is odd under the $\cal CP$ transformation and thus
antisymmetric under the exchange of the electron and positron energies or
momenta. This is born out by explicit calculation, with equation
(\ref{gleichung3.2}) being linear in $\Delta$. The $A_{1}^{+}$ amplitude
is even under exchange of the electron
and positron momenta. Therefore the term $A_{2}^{+} A_{1}^{+ *}$ drops
out in an experiment which symmetrizes over the electrons and positrons.
The remaining interference terms in equation (\ref{gleichung2.1}) are
$\cal CP$-odd. Thus the presence
of an oscillation in the interference region is a clear indication of
$\cal CP$ violation. In the remaining article we estimate each of the
amplitudes, we calculate the magnitude of the effect and demonstrate it
with formulas and several figures.

\section{Estimates for the Amplitudes}

\subsection{\label{directCPv}The direct $\cal CP$-violating amplitude
$B$}
Among the amplitudes, $B$ is the best known in the standard model.
As discussed by several authors, the $B$-amplitude is calculated
according to Fig. \ref{eff-Ham} by means
of an effective Hamiltonian for $\Delta S = 1$ semileptonic transitions
derived by using the operator-product-expansion
$[ 7 - 10 ]$.
One starts at a high energy scale, where the interaction is point-like,
and scales down to low energies by means of the renormalization group
equation. The procedure also includes electromagnetic and strong effects
contained in the Wilson coefficients.
Since the $B$-amplitude is $\cal CP$-violating, it involves the imaginary
parts of the Wilson coefficients, and the dominant terms are
\cite{Donoghue-Holstein-Valencia}
\begin{equation}
Im( C_{7} )\;O_{7},\quad\mbox{and}\quad Im( C_{8} )\;O_{8}
\end{equation}
with
\begin{equation}
O_{7}\;=\;(\overline{s}_{L}\;\gamma_{\mu}\;d_{L})\;(\overline{e}\;
\gamma^{\mu}\;e),\quad\mbox{and}\quad
O_{8}\;=\;(\overline{s}_{L}\;\gamma_{\mu}\;d_{L})\;(\overline{e}\;
\gamma^{\mu}\;\gamma_{5} e).
\end{equation}
These Wilson coefficients receive their main contribution from energy
scales between $m_{t}$ and $m_{c}$ where perturbative QCD is more
reliable. The reduced matrix elements of the operators involve quark
currents between hadronic states and can be related to $K_{l3}$ decays
through an isospin rotation \cite{Donoghue-Holstein-Valencia}.
Neglecting the mass of the electron, the final form of the amplitude is
\begin{equation}
B\;=\;\overline{u}(k_{-}) \slp_{K} \Big[ B^{+} + B^{-} \gamma_{5} \Big]
v(k_{+})
\end{equation}
with
\begin{displaymath}
B^{+}\;=\;i \frac{G_{F}}{\sqrt{2}}\;V_{ud} V_{us}^{*}\;\alpha\;
2 \sqrt{2} f_{+}(s) Im( C_{7} ), \quad
B^{-}\;=\;i \frac{G_{F}}{\sqrt{2}}\;V_{ud} V_{us}^{*}\;\alpha\;
2 \sqrt{2} f_{+}(s) Im( C_{8} ) .
\end{displaymath}
Following the Kobayashi Maskawa parametrization for the quark mixing
matrix we use for the coefficients the values
\cite{Buras-Lautenbacher-Misiak-Muenz}
\begin{equation}
Im( C_{7} ) = - \frac{1}{V_{ud} V_{us}^{*}}\;\mbox{Im}( V_{td}V_{ts}^{*})
\;0.74 \quad \mbox{and} \quad
Im( C_{8} ) = - \frac{1}{V_{ud} V_{us}^{*}}\;\mbox{Im}( V_{td}V_{ts}^{*})
\;( - 0.70 )
\end{equation}
with $m_{t} = 170$ GeV.
The main uncertainty comes from the factor
\begin{equation}
Im( \lambda_{t} )\;=\; Im( V_{td}^{} V_{ts}^{*} )\;=\;
- s_{1} s_{2} s_{3} c_{2} sin \delta
\end{equation}
for which we will allow the range $( 1.0 - 2.0 ) \times 10^{-4}$.
The branching ratio from the $B$-amplitude alone is given by the formula
\begin{equation}
\frac{d \Gamma}{d s d\Delta}\;=\;\frac{1}{512 \pi^{3} m_{K}^{3}}\;
\left[ | B^{+} |^{2} + | B^{-} |^{2} \right]\;\frac{1}{2}\;
\left[ \lambda(s,m_{K}^2,m_{\pi}^{2})  - \Delta^{2} \right] .
\end{equation}
Varying the parameters, we obtain the range
\begin{equation}
BR( K_{L} \rightarrow \pi^{0} e^{+} e^{-} )_{direct}\;=\;
( 2.4 - 9.7 ) \times 10^{-12}.
\end{equation}
Later on we will use the $B$-amplitude as given above with the
corresponding ranges of the parameters in order to study the
development of a pure $K^{0}$ beam.

\subsection{The $\cal CP$-Conserving Amplitude $A_{1}$}
The amplitude $A_{1}$ has the same diagrams ( see Fig. \ref{eff-Ham} )
as the amplitude $B$.
But the approach of using the same effective Hamiltonian as in section
\ref{directCPv} involves the real parts of the Wilson coefficients with
large contributions from regions far below $m_{c}$, where
perturbative QCD is not reliable. For this reason one does not use the
QCD effective Hamiltonian, but resorts to other low energy methods like
chiral perturbation theory.
We define $A_{1}$ through the equation
\begin{equation}
A_{1}\;=\;<\pi^{0}\;e^{+} e^{-}|{ \cal H}_{\gamma}|K_{1} >\;=\;
\overline{u}(k_{-})\;A^{+}_{1} \slp_{K} \; v(k_{+}) ,
\end{equation}
and $A^{+}_{1}$ is given by
\begin{displaymath}
A_{1}^{+}\;=\;\frac{G_{F}}{\sqrt{2}}\;V_{ud} V_{us}^{*} \; g_{8} \;
\frac{\alpha}{\pi} \; 2 \; \left[ w_{+} + \frac{1}{6} \;
\ln{ \frac{m_{\pi}^{2}}{m_{K}^{2}} } \; + 2 \; \phi( s ) \; \right]
\end{displaymath}
with the loop-function $\phi( s )$
\begin{equation}
\phi( s )\;=\;-\frac{4 m_{K}^{2}}{3 s}\;+\;\frac{5}{18}\;+\;\frac{1}{3}\;
\left( \frac{4 m_{K}^{2}}{s}\;-\;1 \right)^{\frac{3}{2}}\;
\arctan\left(1/\sqrt{\frac{4 m_{K}^{2}}{s}\;-\;1}\;\right)
\end{equation}
as calculated in \cite{Ecker-Pich-deRafael}.
Most of the factors here have standard definitions except for $g_{8}$,
which is the coupling constant of the octet of pseudoscalar mesons
and $\omega_{+}$, which is a dimensionless coupling constant. Both have
to be determined experimentally. From the decay $K \rightarrow \pi \pi$
it was found $g_{8} = 5.1$. $\omega_{+}$ was determined from a
$\chi^2$-analysis of the spectrum for the decay
$K^{+} \rightarrow \pi^{+} e^{+} e^{-}$, based on a calculation of the
spectrum in $\chi$PT \cite{Ecker-Pich-deRafael}, including the same set
of parameters. From their data set the BNL E777 group \cite{Alliegrot}
derived a value of
\begin{equation}
w_{+} \; = \; 0.89
\begin{array}{cc}
\scriptstyle + &\scriptstyle 0.24 \\
\scriptstyle - &\scriptstyle 0.14
\end{array} .
\end{equation}
The decay-width from the $A_{1}$ amplitude reads
\begin{equation}
\frac{d \Gamma}{d s d\Delta}\;=\;\frac{1}{512 \pi^{3} m_{K}^{3}}\;
| A^{+}_{1} |^{2}\;\frac{1}{2}\;
\left[ \lambda(s,m_{K}^2,m_{\pi}^{2})  - \Delta^{2} \right] .
\end{equation}
This yields a branching ratio for the decay
\begin{equation}
BR(K_{L} \rightarrow \pi^{0} e^{+} e^{-} ) \; = \;
1.71 \times 10^{- 15} - 1.14 \times 10^{-12}
\end{equation}
if the decay occurs only through the $\cal CP$-violating piece of the
$K^{0}$-$\overline{K}_{0}$ mass-matrix.

In the same way as $B$, the $A_{1}$-amplitude with the corresponding
range of parameters
will be used as an input for the time development of a pure $K^{0}$ beam.

\subsection{The $\cal CP$-conserving amplitude $A_{2}$}
We have already mentioned that we should include the $A_{2}$ amplitude
even though it is of O($\alpha^2 $). Since the decay
$K^{0} \rightarrow \pi^{0} e^{+} e^{-}$ has not yet been observed it was
suggested to study the intermediate decay $K_{L} \rightarrow \pi^{0}
\gamma \gamma$, which has recently been observed, with the branching
ratio \cite{Barr}
\begin{equation}
BR(K_{L} \rightarrow \pi^{0} \gamma \gamma ) \; = \;
( 1.7 \pm 0.3 ) \times 10^{- 7}.
\label{gleichung3.3}
\end{equation}
Starting from these studies, one should couple the two photons to the
final electron-positron pair. This is useful but not very direct, because
some amplitudes contributing to  \\
$K_{L} \rightarrow \gamma \gamma$ are
suppressed in the $K_{2} \rightarrow \pi^{0} e^{+} e^{-}$ amplitude,
being proportional to $m_{e}$ and their contribution being negligible.
As will become clear later on, an amplitude which gives a significant
contribution to the $m_{\gamma \gamma}$-distribution gives a very small
contribution to the semileptonic decay.

The amplitude for the decay $K_{L} \rightarrow \pi^{0} \gamma \gamma$ has
been estimated by two different methods. One method uses a two-component
model developed by Sehgal and collaborators \cite{Sehgal1,Sehgal2}.
The two contributions included are: ($\alpha$) a diagram with a charged
pion loop to which the photons are attached and ($\beta$) vector meson
intermediate states. The second method applies chiral pertubation theory
\cite{Ecker-Pich-deRafael}.
The two approaches differ in several respects, but for the
amplitude which is dominant in our investigation they agree.
This comes about as follows: the decay $K_{L} \rightarrow \pi^{0} \gamma
\gamma$ has several amplitudes, but only one of them is significant for
$A_{2}$. It is fortunate that estimates of this amplitude give similar
results in the two methods. We describe the results of the two-component
model \cite{Sehgal3,Heiliger-Sehgal}.

The amplitude for $K_{L}( p_{K} ) \rightarrow \pi^{0}( p_{\pi})
\gamma( k ) \gamma( k^{'} )$ has the general structure
\begin{eqnarray}
M&=&\epsilon^{\mu} \epsilon^{\nu}\;\Big[ F \; (\;k^{'}_{\mu} k_{\nu}\;
-\;g_{\mu \nu}\;k^{'} \cdot k\;) \nonumber \\
& & \hspace{1cm} + \; G \; (\;g_{\mu \nu}\;k\cdot p_{K}\;k^{'}\cdot p_{K}
\;+\;p_{K \mu} \; p_{K \nu} \; k^{'}\cdot k \label{gleichung3.1} \\
& & \hspace{1.8cm}\;-\;p_{K \mu}\;k_{\nu}\;k^{'}\cdot p_{K}\;
-\;k^{'}_{\mu}\;p_{K \nu}\;k\cdot p_{K}\;) \Big] . \nonumber
\end{eqnarray}
The pion-loop diagrams contribute only through the $F$ amplitude, whose
contribution to $A_{2}$ is proportional to $m_{e}$ and thus small.
This follows by considering the general structure of the loop integral
and the tensor structure of the term that multiplies the $F$ amplitude in
equation (\ref{gleichung3.1}). The vector meson pole diagram contributes
to the amplitude $A_{2}^{+}(s, \Delta)$ defined by
\begin{equation}
A_{2}\;=\;A(K_{2} \rightarrow \pi^{0} e^{+} e^{-} )_{ 2 \gamma} \; = \;
\overline{u}(k_{-})\;A_{2}^{+}\;\slp_{K}\;v(k_{+}) .
\end{equation}
Conservation of $\cal CP$ demands that $A_{2}^{+}(s, \Delta)$ is an odd
function of $\Delta$. Defining \\
$\beta\;=\;\sqrt{ 1 - \frac{4\;m_{e}^{2}}{s} \; }$,
the absorptive part of $A_{2}^{+}(s, \Delta)$ is
\begin{eqnarray}
\mbox{Im}\;A_{2}^{+}&=&\frac{\alpha}{16}\;\frac{G_{eff}}{m_{V}^{2}}\;
\frac{\Delta}{\beta}\;\left[ \frac{2}{3}\;+\;\frac{2}{\beta^{2}}\;-\;
\left( \frac{1}{\beta^{2}}\;-\;\beta^{2}\;\right)\;\frac{1}{\beta}\;
\ln{\frac{1 + \beta}{1 - \beta}}\;\right], \nonumber \\
&=&\frac{\alpha}{16}\;\frac{G_{eff}}{m_{V}^{2}}\;
\frac{8}{3}\;\Delta\hspace{3cm}\mbox{when}\;\beta\;\rightarrow\;1 .
\label{gleichung3.2}
\end{eqnarray}
We notice that $Im A_{2}^{+}$ is an odd function of $\Delta$ and in
addition the limit $\beta \rightarrow 1$ is justified for the decay
$K_{2} \rightarrow \pi^{0} e^{+} e^{-}$. In this limit, equation
(\ref{gleichung3.2}) is in agreement with formula (28) of
\cite{Flynn-Randall2}. The formula above has only one coupling $G_{eff}$
which is chosen in such a way that it reproduces the branching ratio in
equation (\ref{gleichung3.3}).

The dispersive part is calculated with the help of a dispersion relation
\cite{Heiliger-Sehgal}
\begin{equation}
\label{eqn2}
\mbox{Re}\;A_{2}^{+}\;=\;\frac{1}{\pi}\;
\int\limits_{s_{min(\Delta)}}^{\Lambda^{2}\;=\;m_{\rho}^{2}}\;
\frac{ \mbox{Im}\;A_{2}^{+}(\Delta)}{z - s}\;dz
\end{equation}
where the lower limit is given by
\begin{equation}
s_{min}(\Delta)\;=\;4\;m_{e}^{2}\;\left[ 1 -
\frac{\Delta^{2}}{((m_{K} + m_{\pi})^{2} - 4 m_{e}^{2})\;
((m_{K} - m_{\pi})^{2} - 4 m_{e}^{2})}\;\right]^{-1}
\end{equation}
and the upper limit is determined by the heaviest particle considered,
this being the $\rho$ meson.
The differential decay-width reads
\begin{equation}
\frac{d \Gamma}{d s d\Delta}\;=\;\frac{1}{512 \pi^{3} m_{K}^{3}}\;
| A^{+}_{2} |^{2}\;\frac{1}{2}\;
\left[ \lambda(s,m_{K}^2,m_{\pi}^{2}) - \Delta^{2} \right] .
\end{equation}
Inserting the numerical values we obtain a branching ratio dominated by
the vector meson coupling constant
\begin{eqnarray}
Br(K_{L} \rightarrow \pi^{0} e^{+} e^{-} )_{2 \gamma} & = &
4.61 \times 10^{- 12} \left( \frac{G_{eff}\;m_{K}^{2}}{0.25 \times
10^{- 7}}\;\right)^{2}\;( 1 + \rho ) \\
& = & 4.15 \times 10^{- 12} \nonumber \\
\mbox{with} \; \rho & = & \Gamma_{disp} / \Gamma_{abs} \; = \; 1.5 \\
\mbox{and}\; G_{eff}\;m_{K}^{2} & = & 0.15 \times 10^{- 7} . \nonumber
\end{eqnarray}

The second method for calculating the decay $K_{L} \rightarrow \pi^{0}
\gamma \gamma$ is chiral perturbation theory. The authors include
effects of the order $p^{4}$
\cite{Ecker-Pich-deRafael2} and $p^{6}$ \cite{Cohen-Ecker-Pich} in the
momentum expansion of $\chi$PT, as well as vector mesons
\cite{Ecker-Pich-deRafael3}. This enables them to reproduce the observed
decay rate and  spectrum. In this approach there are again two effective
coupling constants which have to be fixed experimentally. Here the
vector meson coupling constant was chosen in such a way so that the
measured decay rate is reproduced.

The calculation of the two-photon-exchange contribution to the decay
$K_{L} \rightarrow \pi^{0} e^{+} e^{-}$ on the basis of the $\chi$PT
prediction is analogous. The branching ratio achieved in this manner
is dominated by the vector meson intermediate state and reads
\begin{eqnarray}
Br(K_{L} \rightarrow \pi^{0} e^{+} e^{-} )_{2 \gamma} & = &
1.8 \times 10^{- 12} ( 1 + \rho ) \\
& = & 4.5 \times 10^{- 12} \nonumber
\end{eqnarray}
with the same $\rho$.
In the numerical analysis of section \ref{numerik} we can choose any of
the two calculations for $K_{2} \rightarrow \pi^{0} e^{+} e^{-}
|_{2 \gamma}$, because the term relevant for our purpose is practically
the same. In addition, we shall demonstrate that in experiments which
average over the $e^{+}$ and $e^{-}$ momenta, this $A_{2}$ amplitude
drops out in the interference region.

\section{\label{numerik} Numerical results}
With the amplitudes developed and the general equation
(\ref{gleichung2.1}) we calculate the time evolution for the decay
$K^{0} \rightarrow \pi^{0} e^{+} e^{-}$. As already discussed, there are
three regions of physical interest.
\begin{itemize}
\item[1)] The $K_{S}$-region, where the $\cal CP$-conserving amplitude
contributes to the decay width. A measurement in this region will
determine parameters of the $K_{S}$-decay like $\omega_{+}$.
\item[2)] The $K_{L}$-decay region, which has been studied in several
articles interested in $\cal CP$ phenomena ( see the review
\cite{Winstein-Wolfenstein,Rosner} ). The relevant amplitude in this
case is $| B + A_{2} + \epsilon A_{1} |^{2}$, which gives several terms.
The interference term $Re( A_{2} B^{+} + \epsilon A_{1} A_{2} )$ is odd
in $\Delta$ and one could define an asymmetry in $\Delta$ in order to
extract this term. The sum of the absolute values squared is even in
$\Delta$, and we will need precise measurements of the amplitudes in
order to observe an excess of events.
\item[3)] More interesting is the interference region occuring in the
time interval $( 8 - 9 )\;\cdot\tau_{K_{S}}$. This term has an
oscillatory behavior. The term $A_{2} A_{1}^{+}$ is a linear function in
$\Delta$ and drops out when we average over the electron positron pair.
The remaining terms $B A_{1}^{+}$ and $\epsilon | A_{1} |^{2}$ are both
$\cal CP$-violating. Thus the appearance of an oscillation in the
interference region gives evidence for $\cal CP$ violation.

We have studied this phenomena numerically and show the effect in several
figures. Fig. (2) and Fig. (3) show the branching ratio as a function of
time for two different time scales. We note that an oscillation is
evident. We use $\omega_{+} = 0.89$ and three values for
$Im \lambda_{t} = 1.0 \times 10^{-4}, 1.5 \times 10^{-4}\;\mbox{and}\;2.0
\times 10^{-4}$. We plotted the same curves in figures (4) and (5) where
the $\cal CP$-violating terms $B A_{1}^{+}$ and $\epsilon | A_{1} |^{2}$
are set equal to zero. We noticed that the curves to be compared are
different. In the interference region there is a clear oscillation and
for very long times the curves which contain the $\cal CP$ amplitudes lie
above the curve without the $\cal CP$-violating terms. For the latter
region experiments studying $K_{L}$-decays need a precise measurement of
magnitudes in order to establish a signal. The branching ratio is in the
range $(3 - 5) \times 10^{-12}$. In contrast to this situation, the
oscillation in the interference region is unambiguous.
A comparison of the magnitudes of the contributing terms shows that
$B A_{1}^{+}$ dominates over $\epsilon | A_{1} |^{2}$. Only if we choose
$\omega_{+}$ at the upper bound ($\omega_{+} = 1.13$) the two
contributions are of comparable size. But the direct $\cal CP$-violating
term is still larger by $(6 - 113) \%$ for $Im(\lambda_{t}) = (1.0 - 2.0)
\times 10^{-4}$.
\end{itemize}
We conclude that an experiment searching for a branching ratio down to
$10^{-12}$ and sensitive to the time development of the decay can observe
$\cal CP$ violation as an oscillation in the interference region. The
experiment does not require a measurement of the $e^{+} e^{-}$ energy
asymmetry.

\vspace{1cm}

\newpage

\centerline{ \large  Acknowledgements }

\vspace{0.5cm}

We wish to thank Drs. Y.L. Wu and P. Overmann for discussions. One of us
(EAP) thanks Dr. L. Littenberg for a motivating discussion on this topic.

\newpage

{\large \bf Figure Captions }\\
\begin{itemize}
\item[Fig. \ref{eff-Ham}] Short distance contributions to the decay
$K \rightarrow \pi^{0} e^{+} e^{-}$
\item[Fig. 2] Time development of the partial branching ratio
$\Gamma( K^{0} \rightarrow \pi^{0} e^{+} e^{-})(t)/\Gamma( K_{L}
\rightarrow all )$ for the \\
time interval $( 6 - 20 )\;\cdot\;\tau_{K_{S}}$.
\item[Fig. 3] Same as in fig. 2 for a larger time interval.
\item[Fig. 4] The same curves as in fig. 2 together with the modified
decay rate in which the $\cal CP$-violating terms are set to zero.
\item[Fig. 5] The same curves as in fig. 3 together with the modified
decay rate when the $\cal CP$-violating terms are set to zero.
\end{itemize}

\newpage

\epsfverbosetrue
\epsfbox{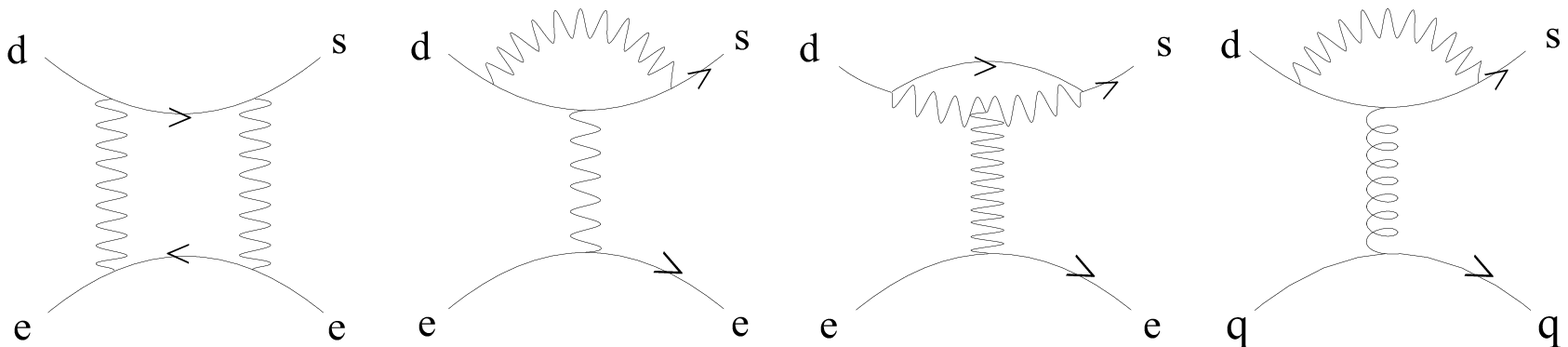}

\begin{figure}
\caption{\label{eff-Ham}Short distance contributions to the decay
$K \rightarrow \pi^{0} e^{+} e^{-}$}
\hfill
\end{figure}

\newpage

\epsfverbosetrue
\epsfbox{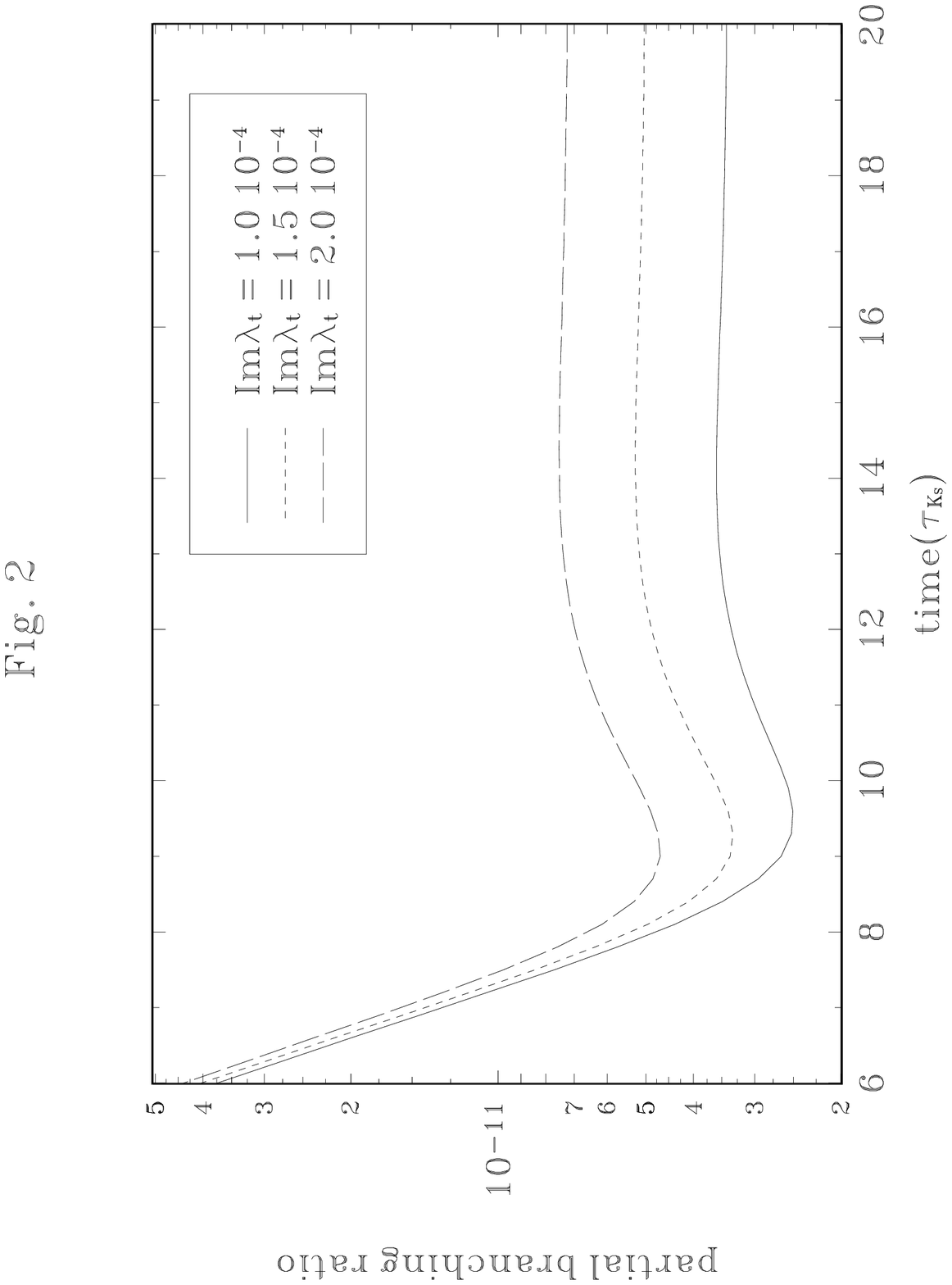}

\begin{figure}
\caption{Time development of the decay $\Gamma( K^{0} \rightarrow
\pi^{0} e^{+} e^{-}/\Gamma( K^{0} \rightarrow all )$ for the
time interval $( 6 - 20 )\;\tau_{K_{S}}$ }
\end{figure}

\newpage

\epsfverbosetrue
\epsfbox{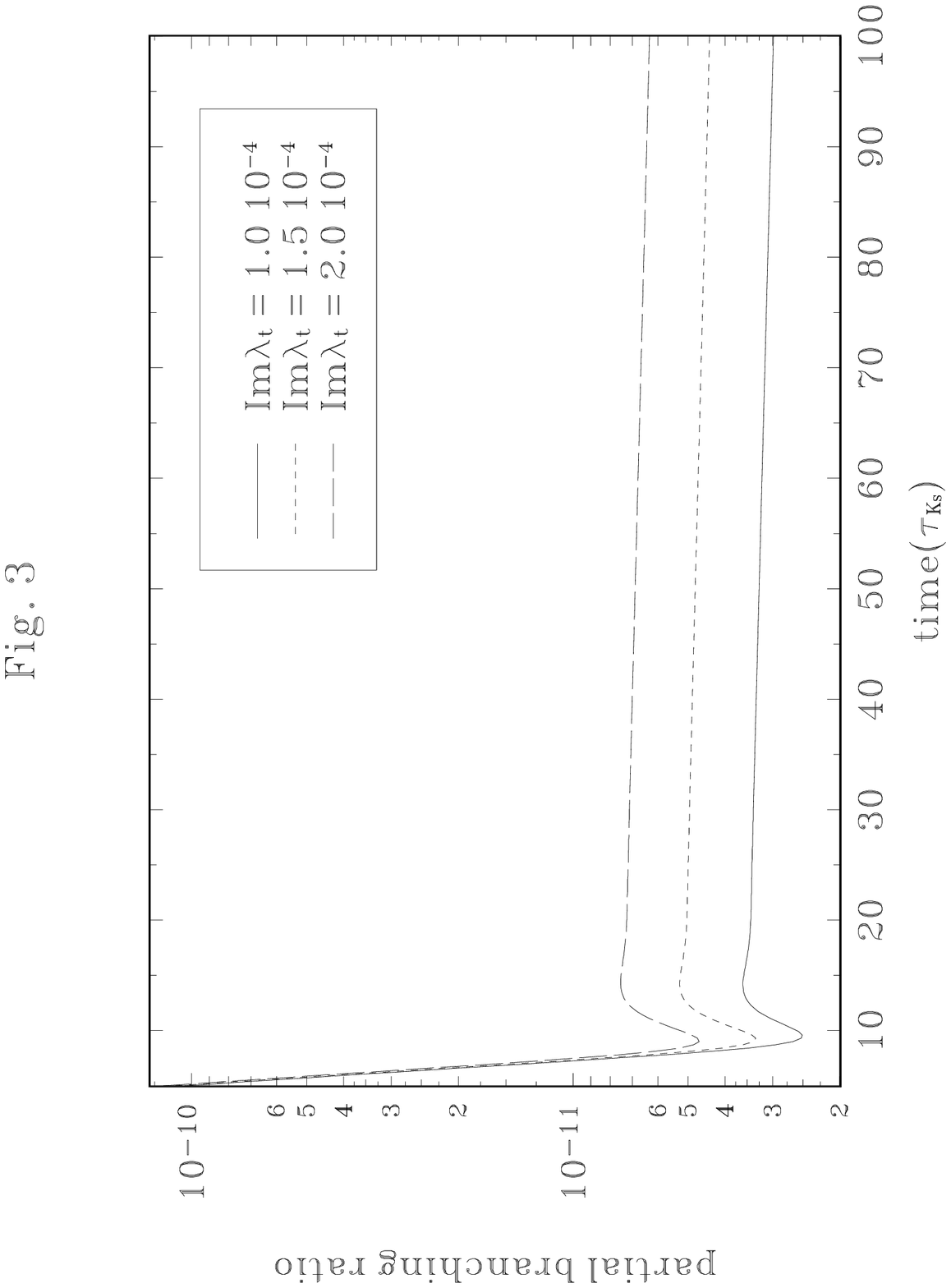}

\begin{figure}
\caption{Same as in fig. 2 for a longer time interval.}
\end{figure}

\newpage

\epsfverbosetrue
\epsfbox{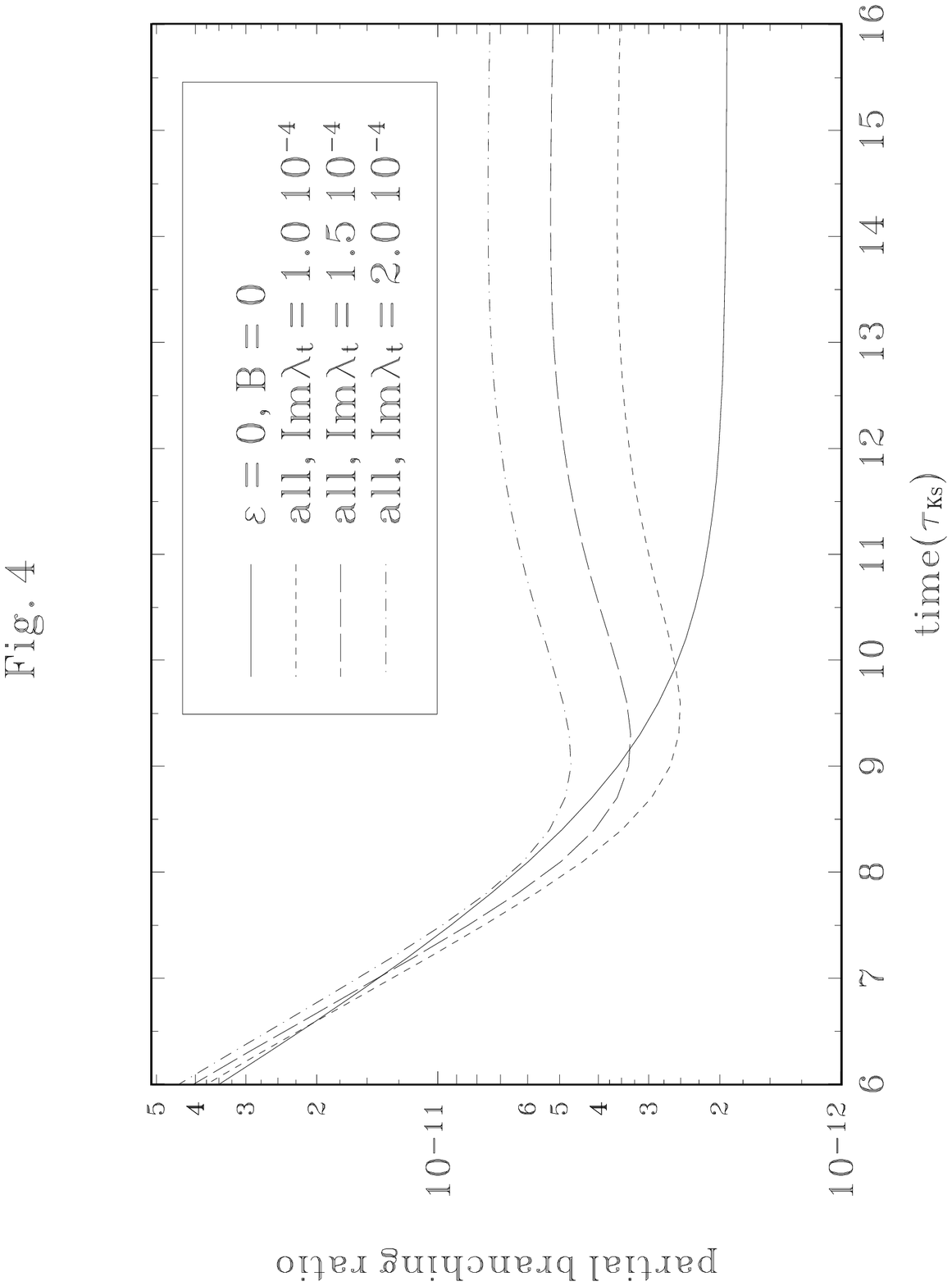}

\begin{figure}
\caption{The same curves as in fig. 2 together with the decay when the
$\cal CP$-violating terms are set to zero.}
\end{figure}

\newpage

\epsfverbosetrue
\epsfbox{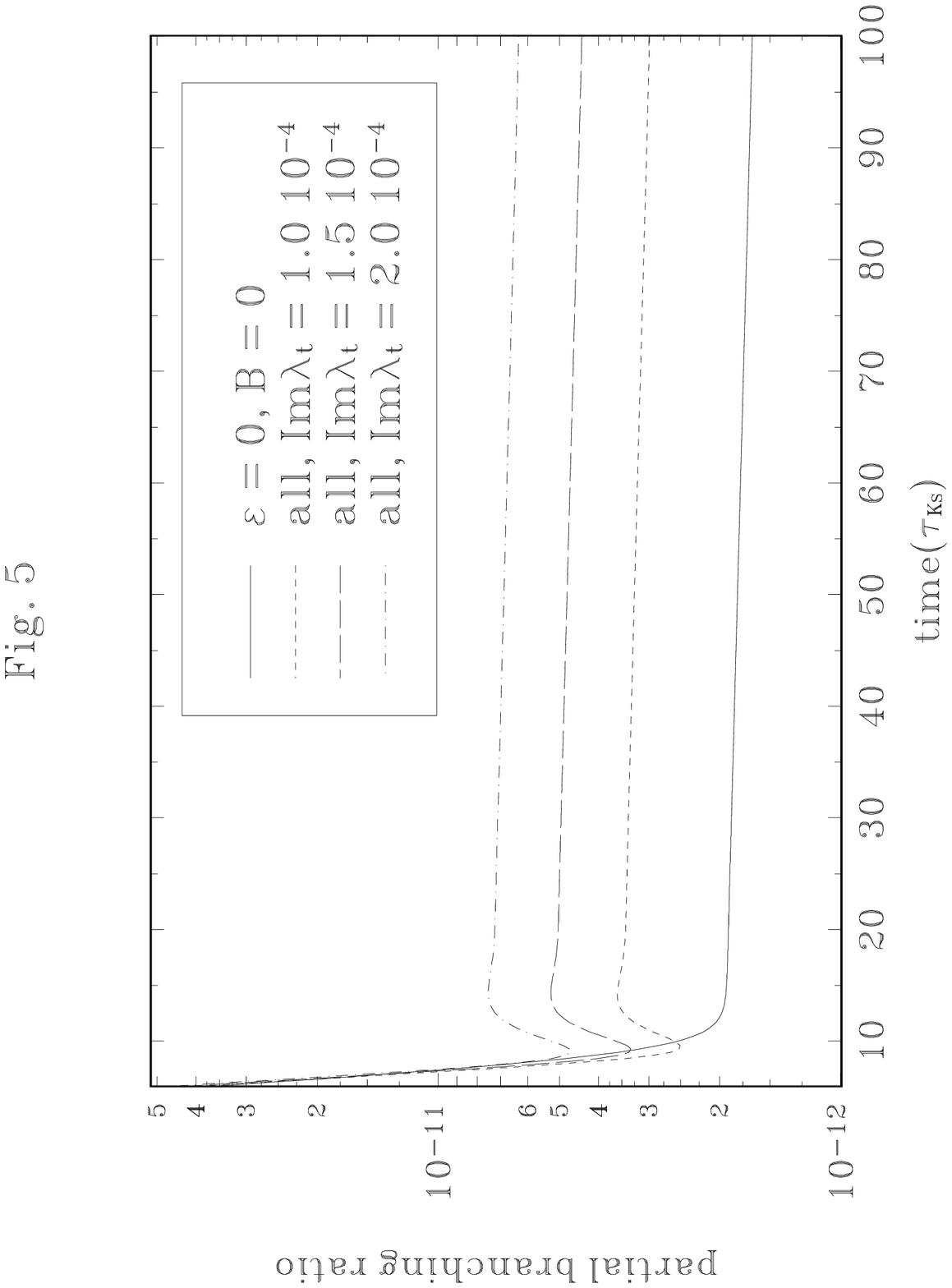}

\begin{figure}
\caption{The same curves as in fig. 3 together with the decay when the
$\cal CP$-violating terms are set to zero.}
\end{figure}

\end{document}